\documentclass[aps,prl,groupedaddress,fleqn,twocolumn]{revtex4}
\pdfoutput=1
\usepackage{graphicx}
\usepackage{amsmath}
\usepackage{gensymb}
\usepackage{multirow, array}
\usepackage{adjustbox}

\begin{document}
\title{The nature of collective excitations and their crossover at extreme supercritical conditions}
\author{L. Wang$^{1}$}
\author{C. Yang$^{1,2}$}
\author{M. T. Dove$^{1}$}
\author{A. V. Mokshin$^{3,4}$}
\author{V. V. Brazhkin$^{5}$}
\author{K. Trachenko$^{1}$}
\address{$^1$ School of Physics and Astronomy, Queen Mary University of London, Mile End Road, London, E1 4NS, UK}
\address{$^2$ Shanghai Jiao Tong University, China}
\address{$^3$ Landau Institute for Theoretical Physics, Russian Academy of Sciences, 142432 Chernogolovka, Russia}
\address{$^4$ Institute of Physics, Kazan Federal University, 420008 Kazan, Russia}
\address{$^5$ Institute for High Pressure Physics, RAS, 108840, Moscow, Russia}
\address{*Correspondence to ling.wang@qmul.ac.uk}

\begin{abstract}
Physical properties of an interacting system are governed by collective excitations, but their nature at extreme supercritical conditions is unknown. Here, we present direct evidence for propagating solid-like longitudinal phonon-like excitations with wavelengths extending to interatomic separations deep in the supercritical state at temperatures up to 3,300 times the critical temperature. We observe that the crossover of dispersion curves develops at $k$ points reducing with temperature. We interpret this effect as the crossover from the collective phonon to the collisional mean-free path regime of particle dynamics and find that the crossover points are close to both the inverse of the shortest available wavelength in the system and to the particle mean free path inferred from experiments and theory. Notably, both the shortest wavelength and mean free path scale with temperature with the same power law, lending further support to our findings.
\end{abstract}

\maketitle

\section{Introduction}

Any system with interacting agents or particles is capable of supporting waves. The hydrodynamic approximation to liquids and gases and elastic approach to solids, collectively known as the continuum approximation, is widely used in many areas of physics. It predicts propagating long-wavelength longitudinal density fluctuations, sound. As useful as it is, the continuum approximation does not describe many important properties: for example, the solid state theory and in particular thermodynamics of solids relies on the atomistic description. The atomistic description relies, in turn, on the knowledge that the range of wavelengths $\lambda$ in solids varies from approximately system size to the shortest interatomic separation $a$ on the order of Angstroms where atomistic effects operate.

Theoretical understanding of waves in liquids remains challenging, particularly in view that widely-used perturbation approaches do not apply \cite{landau}. Until fairly recently, this difficulty extended to experimental studies as well. However, the deployment of next-generation synchrotron sources enabled physicists to make the same assertion about waves in liquids as in solids: phonon-like excitations in liquids extend up to the shortest interatomic separation as they do in solids \cite{burkel,pilgrim2,rec-review,hoso,hoso3,mon-na,mon-ga,sn}. This solid-like property of liquids is remarkable because traditionally, liquids have been studied in the hydrodynamic approximation $\lambda\gg a$ only \cite{hydro}. It suggests that liquids are amenable to understanding using solid-like concepts and their implications \cite{ropp}.

The recent evidence of non-hydrodynamic solid-like waves in liquids (waves with large $k$ vectors) \cite{burkel,pilgrim2,rec-review,hoso,hoso3,mon-na,mon-ga,sn,ropp} prompts an intriguing question of whether these waves can extend to the third state of matter, gases, and thus be common to {\it all three states}. In familiar gases at ambient conditions, this would appear impossible because the particle mean free path $l$ is much larger than $a$, implying that the system can not support solid-like wavelengths close to $a$. $l$ can be decreased by increasing pressure or lowering the temperature but this results in the first-order gas-liquid transition well before $l$ approaches Angstroms. However, the supercritical state where no gas--liquid phase transition intervenes, offers more flexibility: one can change the density {\it continuously}, from gas-like to liquid-like values. Therefore, studying the supercritical state offers an intriguing possibility to see whether the longitudinal wave in the gas-like state (traditionally known as sound wave) can support ever-decreasing wavelengths up to the shortest solid-like atomistic lengths where $\lambda$ becomes comparable to Angstroms. This would imply new unanticipated properties of waves that gas-like states can support.

The theory of dilute gases is tractable due to the concept of mean free path (MFP), based on the idea that molecules mostly move freely between binary collisions. The theory gives specific predictions for important system properties such as temperature dependence of viscosity and thermal conductivity \cite{chapman}. For dense gases and fluids (loosely defined as systems where intermolecular distances are comparable to molecular sizes) at moderate temperature, this approach does not apply because a molecule continuously moves in the field of forces of others. This involves three and higher-order encounters, and treating these is of considerable difficulty. As a result, common gas theories are unable to describe important properties of dense gases such as temperature dependence of viscosity \cite{chapman}.

The division into dilute and dense gases has traditionally been done at moderate pressure and temperature \cite{chapman}. However, experiments have been increasingly probing the matter at extreme conditions including deeply supercritical ones, calling for new theories to be developed \cite{deben}. Indeed, the supercritical state has remained poorly understood in general, despite the wide deployment of supercritical fluids in important industrial applications \cite{deben}. This has stimulated our interest in a hitherto unexplored question: what kind of collective excitations can exist in the supercritical state of matter?

Central to our proposal is that both temperature and density (rather than density alone as traditionally assumed) are important for the concept of the mean free path to emerge. Indeed, let us consider a supercritical fluid at density close to, for example, water density at the triple point where the system would be characterized as ``dense'' as opposed to ``dilute'' in the traditional classification. If this system is at very high supercritical temperature, a molecule, even though it moves in the field of others, has enough energy to move with little deflection for a certain distance. More specifically, the high energy of the particle results in small deflection angles and small momentum transfer in the collision integrals featuring in transport properties \cite{chapman}. This implies that the concept of the mean free path $l$ emerges in dense systems provided the particle energy is high, and is not limited to dilute gases as assumed previously. In the somewhat crude picture of the mean free path in the system with particles of size $a$, the increase of $l\propto\frac{1}{a}$ corresponds to the decrease of effective $a$ with temperature.

We can therefore develop and use theoretical predictions of dynamical and thermodynamic properties in dense and hot supercritical fluids in the MFP regime. Testing this idea constitutes one of the general aims of this paper.

Apart from the general question of the nature of collective excitations at deep supercritical conditions, we consider the important implication of the MFP regime for the wave propagation. In the MFP regime, the system cannot support an oscillatory motion with wavelengths shorter than the mean free path $l$. We therefore predict that supercritical dispersion curves should undergo a {\it crossover} from the phonon regime at small $k$ vectors to the MFP regime where no phonons exist with wavelengths shorter than $l$.

In this paper, we perform extensive molecular dynamics (MD) simulations and obtain direct evidence for propagating solid-like longitudinal modes with short wavelengths deep in the supercritical state at temperatures up to 3,300 times the critical temperature. We subsequently observe that the crossover of dispersion curves develops at $k$ points reducing with temperature and interpret this effect as the crossover from the phonon to the MFP regime on the basis of (a) closeness of the crossover points to those corresponding to the shortest available wavelength in the system; (b) closeness of the crossover points to those corresponding to the particle mean free path; and (c) the same power-law temperature dependence of the shortest wavelength and the mean free path.

We note that traditionally, the supercritical state was thought to disallow any difference between a gas and a liquid \cite{deben}. We recently proposed that the supercritical state can be separated into the gas-like and liquid-like properties by the Frenkel line (FL) \cite{pre,prl,phystoday}. Above the line, particle dynamics are purely diffusive as in gases. In contrast, particle dynamics below the line have two components of motion: solid-like oscillatory motion and diffusive jump motion which enables the system to flow. This regime of particle dynamics below the FL is similar to that in low-temperature liquids, including the viscous liquids in the glass transformation range \cite{pastore1,pastore2}.

Different regimes of particle dynamics above and below the FL have important implications for the ability of the supercritical system to support collective modes. Below the FL, the supercritical system supports both longitudinal and transverse modes \cite{yang,jpcm,jpcm1}. However, temperature increase results in shrinking the range of $k$-points at which transverse operate \cite{yang}. At the FL, transverse modes disappear from the system spectrum, and only one longitudinal mode remains propagating above the FL \cite{jpcm,jpcm1}. Hence the Frenkel line provides a useful guide as to where on the supercritical phase diagram we expect to find a gas-like state with the longitudinal mode only. In this paper we consider the conditions well above the FL in order to study the evolution of the longitudinal mode at deep supercritical conditions.

\section{Results and discussion}

We have performed extensive molecular dynamics (MD) simulations \cite{md} of supercritical liquid Ar \cite{ling} using the Lennard-Jones (LJ) potential and constant volume and energy ensemble. The simulated temperature starts from 500 K (just above the FL temperature) and increases to very high temperatures deep in the supercritical state up to 500,000 K, corresponding to over 3,000 times the critical temperature (Ar critical temperature is 151 K). The simulated densities are $0.4$, $0.65$ and $0.8$ g/cm$^3$. We simulated 8,000 atoms in the system. The time step varies between 1 fs at low temperature and 0.1 fs at high temperature to account for faster dynamics.

The reason for simulating high temperature (the highest temperature is in excess of the Ar ionization energy) is that, as discussed below, $l$ is a slowly-varying function of temperature at constant density. This is due to two competing mechanisms: on one hand temperature increases $l$ but on the other hand the buildup of pressure with temperature at constant density decreases $l$.

\begin{figure}
\begin{center}
{\scalebox{0.5}{\includegraphics{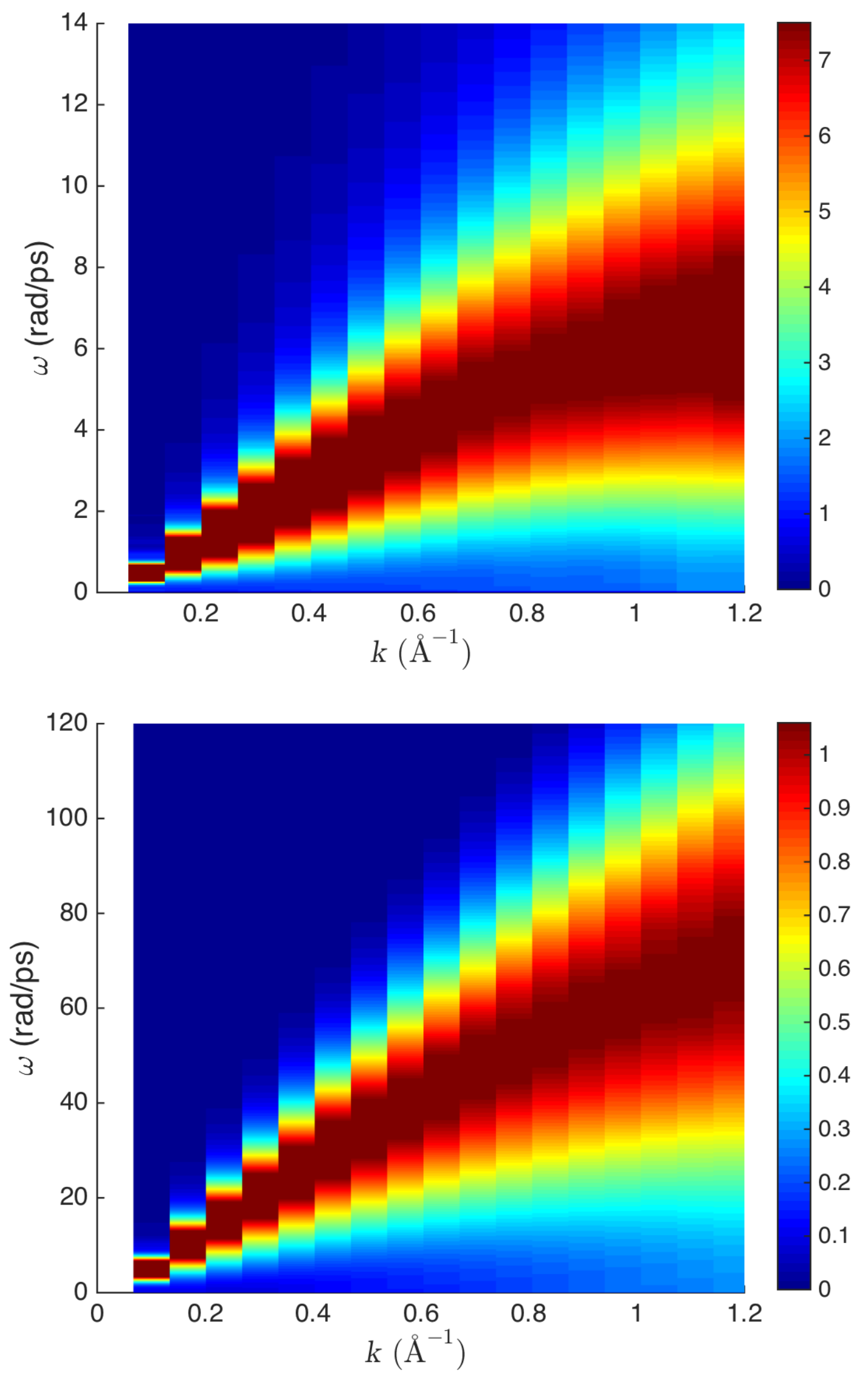}}}
\end{center}
\caption{Intensity map of $C_L( k, \omega)$ for supercritical Ar at (top panel) 500 K and (bottom panel) 100,000 K. The maximal intensity corresponds to the middle points of the dark red areas and reduces away from them.}
\label{intensity}
\end{figure}

We have calculated the longitudinal current correlation functions \cite{Balucani} in order to address the evolution of longitudinal collective modes directly:
\begin{equation}
 C_L (k,t) = (k^2/N) \langle{J_x(\textbf k,t) \cdot (J_x(- \textbf k,0)}\rangle
\label{cl}
\end{equation}
\noindent where $J(\textbf k,t)=\sideset{}{_{j=1}^N}\sum \textbf v_j(t) e^{-i \textbf k \cdot \textbf r_j(t)} $ is the longitudinal current, $N$ is the number of particles, $\textbf v$ is the particle velocity, and the wave vector $\textbf k$ is along the $x$ axis. $k$-points were sampled as $k=2\pi n/L$, where $L$ is the system size.

The spectra of longitudinal currents are calculated by the Fourier transform of $C_L(k,t)$. In order to reduce the noise in calculating the spectra, we have repeated our simulations 20 times for each temperature by using different starting velocities and then we averaged the current results \cite{yang}. Examples of intensity maps of $\tilde C_L(k, \omega)$ are shown in Fig. \ref{intensity}.

We observe that the mode frequency increases with $k$ up to large k-points corresponding to wavelengths comparable to interatomic separations as is the case in solids. In fact, the spectra look remarkably similar to those in solids. The similarity of liquid and solid phonon spectra was noted earlier at low temperature \cite{monaco}. Fig. \ref{intensity} shows that this similarity extends to very high temperature deep into the supercritical state.

\begin{figure}
\begin{center}
{\scalebox{0.33}{\includegraphics{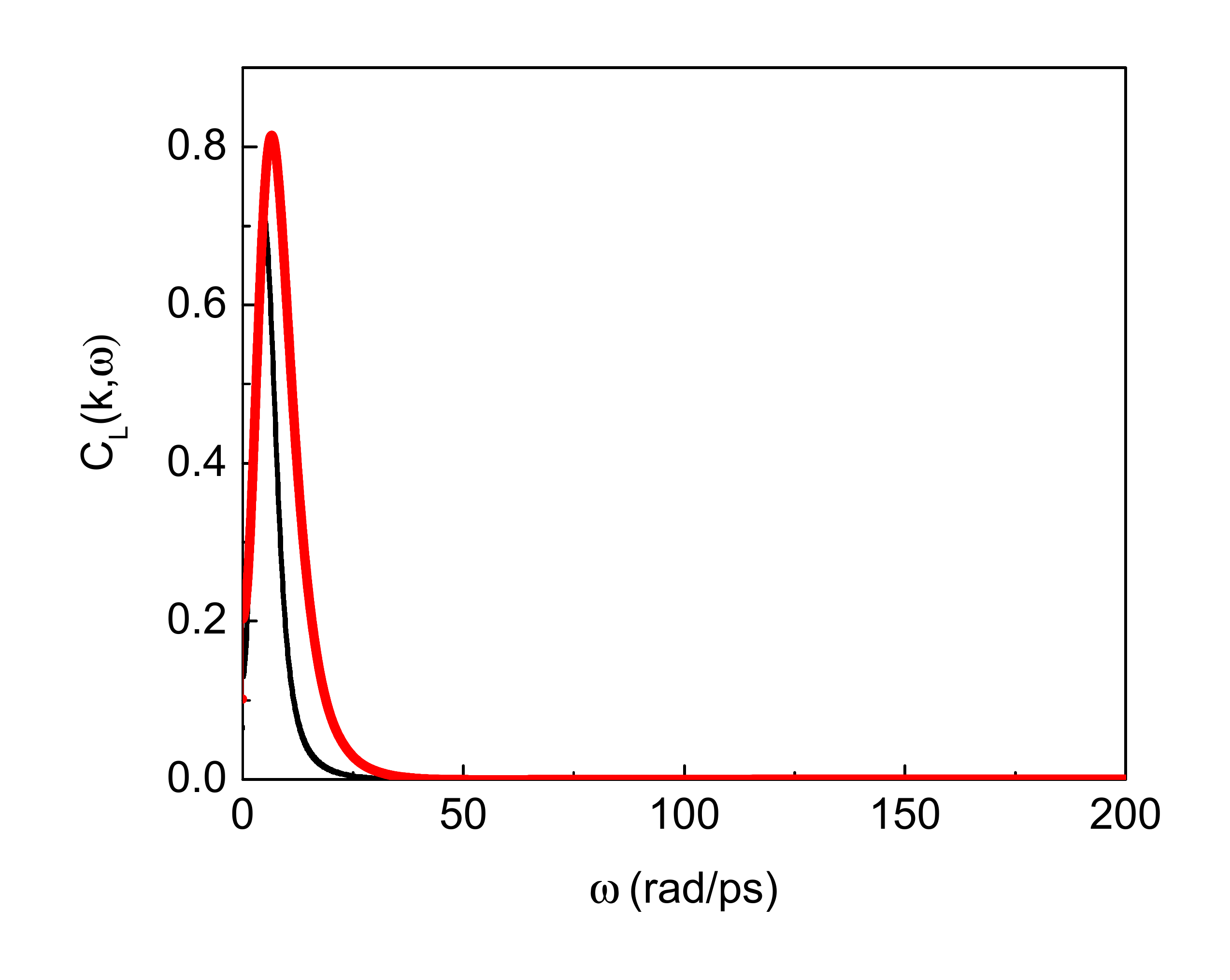}}}
\end{center}
\caption{Examples of $C_L(k,\omega)$ calculated at density 0.65 g/ml and $T=500$ K and shown at $k=0.67$ \AA$^{-1}$ (black) and $k=1.34$ \AA$^{-1}$ (red).}
\label{maxima}
\end{figure}

The frequency at which the intensity is maximal corresponds to the mode frequency at the corresponding $k$-point. We show examples of $\tilde C_L(k, \omega)$ at different $k$ in Fig. \ref{maxima}. The ratio of the peak width at half-height $\Gamma$ to the peak frequency $\omega$ is 1.2-1.5 for the intensity peaks considered in this work, including in the examples shown in Fig. \ref{maxima}. This is consistent with the experimental findings reporting propagating modes in liquids (see, e.g., Ref. \cite{giordano-PRB}). The calculated $\frac{\Gamma}{\omega}$ satisfies the condition $\frac{\Gamma}{\omega}<\sqrt{3}$ for propagating modes derived theoretically \cite{stas}.

\begin{figure*}
\begin{center}
{\scalebox{0.8}{\includegraphics{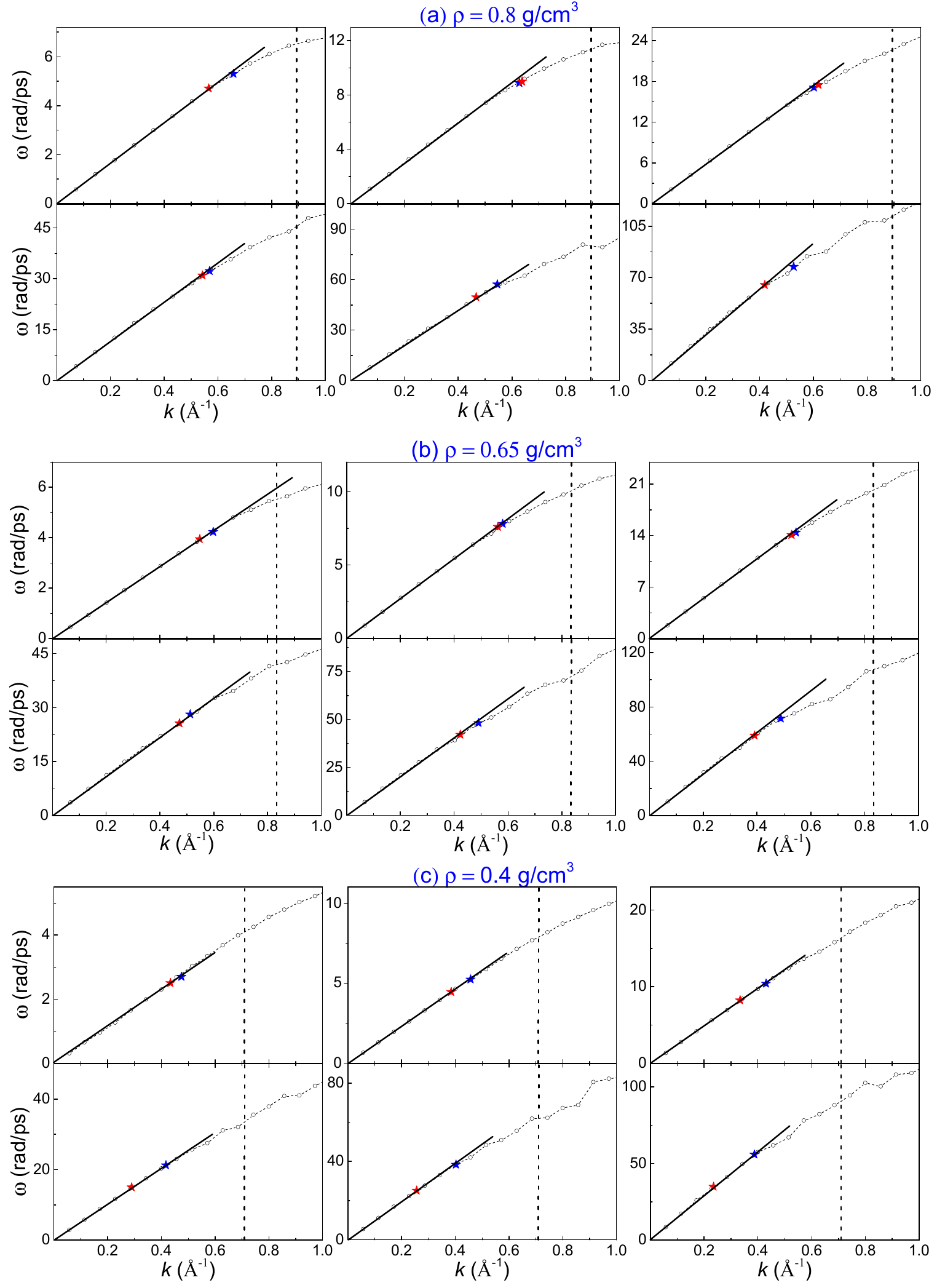}}}
\end{center}
\caption{Phonon dispersion curves of supercritical Ar at density $\rho_1=0.8$ g/cm$^3$ (top panel), $\rho_2=0.65$ g/cm$^3$ (middle panel) and $\rho_3=0.4$ g/cm$^3$ (bottom panel). The corresponding temperatures are (from left to right, top to bottom of each density) are 500 K, 2000 K, 10000 K, 50000 K, 200000 K and 500000 K, respectively. The blue star points are $k^\star$ calculated by Eq.(\ref{lambda}). The red star points are $k$-points calculated as $\frac{\pi}{l}$, where the mean-free path $l$ is evaluated from experimental gas-like viscosity. The dashed vertical lines mark the zone boundary for an isotropic system calculated in the spherical Debye approximation as $k_{ZB}=\left(6\pi^2 \frac{N}{V}\right)^\frac{1}{3}$ \cite{landau}. The straight lines starting from (0,0) guide the eye showing the linear slope.}
\label{dispersion}
\end{figure*}

\begin{figure}
\begin{center}
{\scalebox{0.33}{\includegraphics{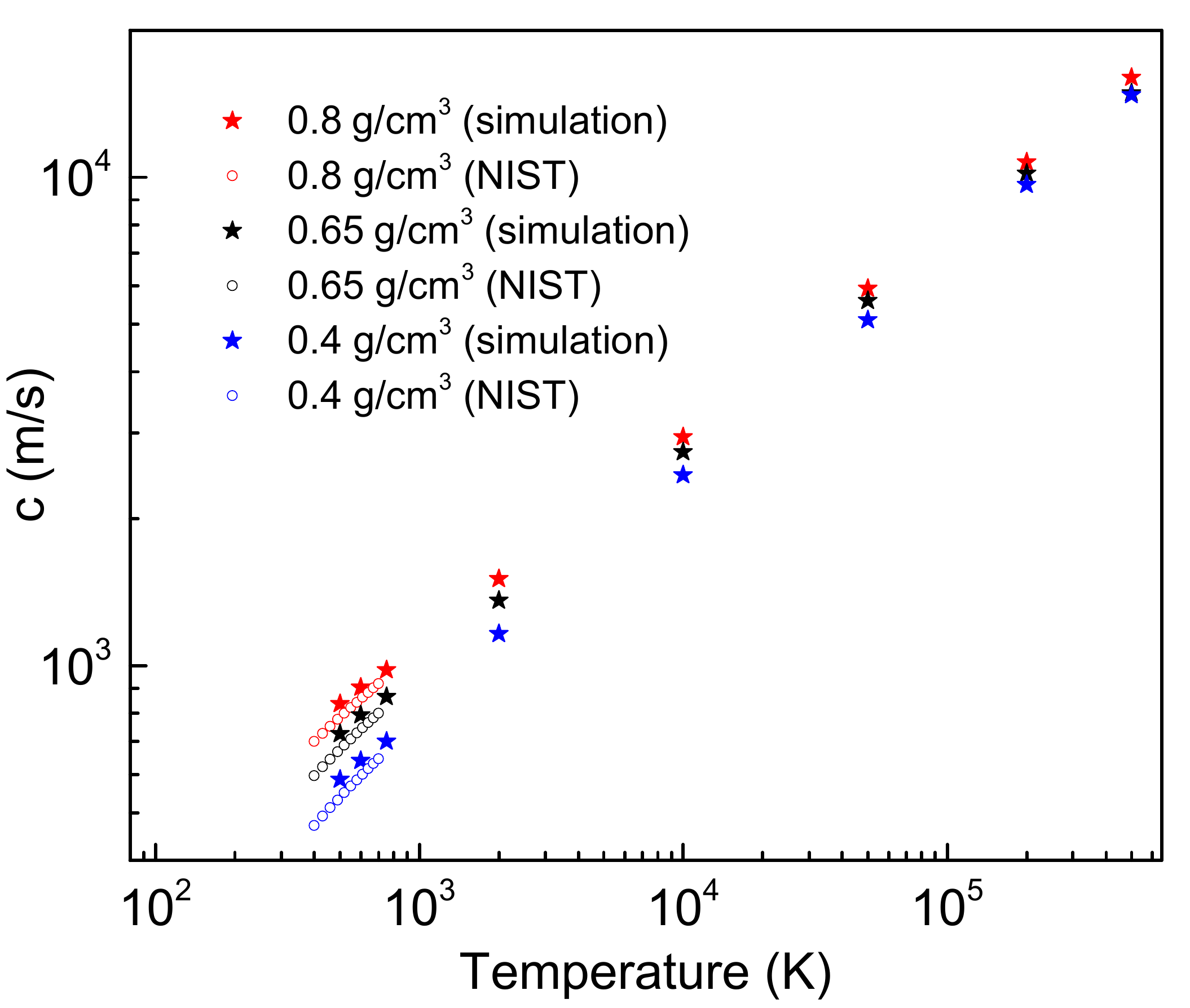}}}
\end{center}
\caption{The calculated speed of sound (stars) and the experimental \cite{NIST} speed of sound (open symbols) at three simulated densities.}
\label{speed}
\end{figure}

We now focus on detailed examination of longitudinal dispersion curves and their change with temperature. The maxima of Fourier transforms $\tilde C_L(k, \omega)$ give the frequencies of longitudinal excitations  \cite{Balucani}. The resulting dispersion curves are shown in Fig. \ref{dispersion}. The slope of the linear part of the dispersion curve at small $k$ gives the speed of sound, $c$, which can be compared with experiments probing the speed of sound at small $k$ and $\omega$. As follows from Fig. \ref{speed}, the calculated $c$ agrees with the experimental $c$ available at low temperature \cite{NIST} well. The increase of $c$ with temperature is due to large pressure buildup in the system at constant density. For example, the pressure increases to 46 GPa, 80 GPa and 103 GPa at the highest simulated temperature at three simulated densities ($0.4$, $0.65$ and $0.8$ g/cm$^3$).

We observe that the regime of collective phonon excitations starts to deviate from linearity in Fig. \ref{dispersion}. This crossover takes place at temperatures well above the FL. This is expected because, as discussed above, crossing the FL from below corresponds to the regime of diffusive particle motion where the concept of the MFP applies.

Importantly, the crossover of dispersion curves takes place at $k$ points which decrease with temperature. This behavior is consistent with our hypothesis above: in the MFP regime above the FL where atoms move freely up to distance $l$ on average, the increase of $l$ with temperature implies that the shortest wavelength of the longitudinal phonons increases, resulting in the decrease of their largest $k$ points. We note that the MFP regime (and the corresponding slope of $\omega$ vs $k$) is intermediate between the sound regime at small $k$ and free-particle regime at large $k$ discussed later.

We support this interpretation with three quantitative estimates. We start with evaluating $\lambda$ and note that one straightforward evaluation involves a relationship between $\lambda$ and specific heat $c_v$. Indeed, free motion of atoms up to distances comparable to $l$ in the MFP regime contributes only to kinetic energy but not to the potential term. Then, the energy of the system with one longitudinal mode only is the sum of the kinetic term $\frac{3}{2}NT$ ($k_{\rm B}=1$) and the potential energy of the longitudinal mode with wavelengths longer than the shortest wavelength in the system $\lambda=l$. Using the Debye model, this gives the energy as \cite{natcom}:

\begin{equation}
E=\frac{3}{2}NT+\frac{1}{2}NT\frac{a^3}{\lambda^3}
\label{energy}
\end{equation}

\noindent where $a$ is interatomic separation.

Eq. (\ref{energy}) gives specific heat $c_v=\frac{3}{2}+\frac{1}{2}\frac{a^3}{\lambda^3}$ (the variation of $\lambda$ with temperature is neglected due to the very slow increase of $l(T)$ as discussed below). This gives $\lambda$ as

\begin{equation}
\lambda=\frac{a}{(2c_v-3)^{1/3}}
\label{lambda}
\end{equation}

\begin{figure}
\begin{center}
{\scalebox{0.33}{\includegraphics{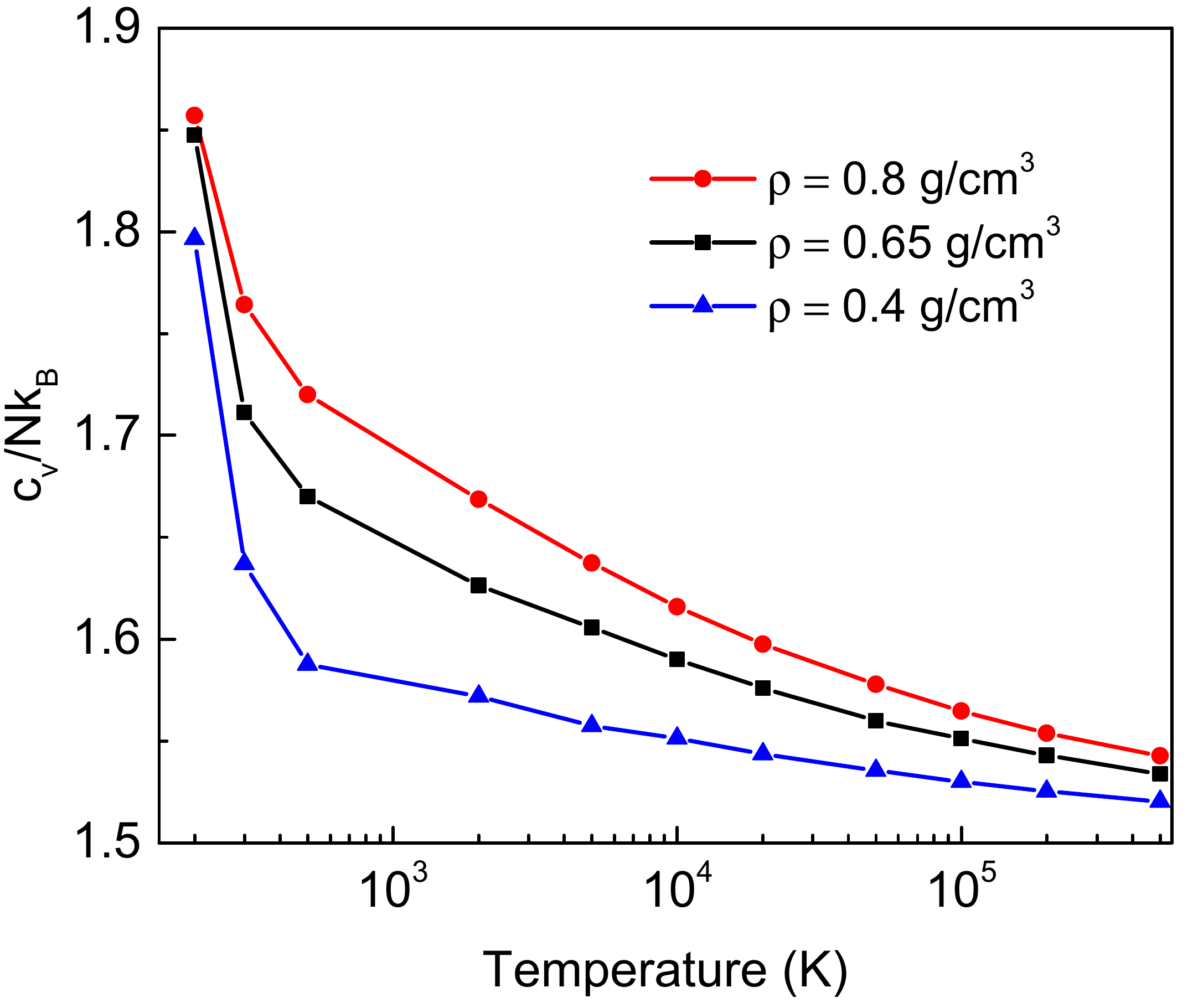}}}
\end{center}
\caption{Specific heat as a function of temperature for three simulated densities.}
\label{cv}
\end{figure}

Eq. (\ref{lambda}) applies to the regime above the Frenkel line where $c_v<2$ \cite{pre,prl,phystoday} (under this condition, Eq. (\ref{lambda}) implies $\lambda>a$ as required). We have calculated $c_v=\frac{1}{N}\frac{dE}{dT}$ in a wide temperature range corresponding to the decrease of $c_v$ from about 2 to its gas-like value $\frac{3}{2}$ (see Fig. \ref{cv}). We observe $c_v$ decreases very slowly at high temperature. According to Eq. (\ref{lambda}), $\lambda$ is predicted to vary similarly slowly at high temperature. This will be discussed later in the paper.

Using the calculated $c_v$, we have evaluated the shortest wavelength $\lambda$ using Eq. (\ref{lambda}). Some care is needed to find the corresponding $k$. In Eq. (\ref{energy}), the shortest $\lambda$ is assumed to be $a$, corresponding to $E=2NT$ and $c_v=2$ \cite{natcom}. For specific evaluations of $\lambda$, we recall that the shortest wavelength in the system with the shortest length scale $a$ is $2a$, therefore $k$ at the crossover from the sound to the MFP regime, $k^\star$, is $k^\star=\frac{2\pi}{2\lambda}=\frac{\pi}{a}\left(2c_v-3\right)^\frac{1}{3}$. We plot $k^\star$ in Fig. \ref{dispersion} as blue stars and observe that in most cases (particularly at higher density) $k^\star$ lies close to the deviation of the sound regime from linearity. This constitutes our {\it first} quantitative evidence in support to our hypothesis that the crossover of dispersion curves is related to the shortest wavelength available in the system.

\begin{figure}
\begin{center}
{\scalebox{0.33}{\includegraphics{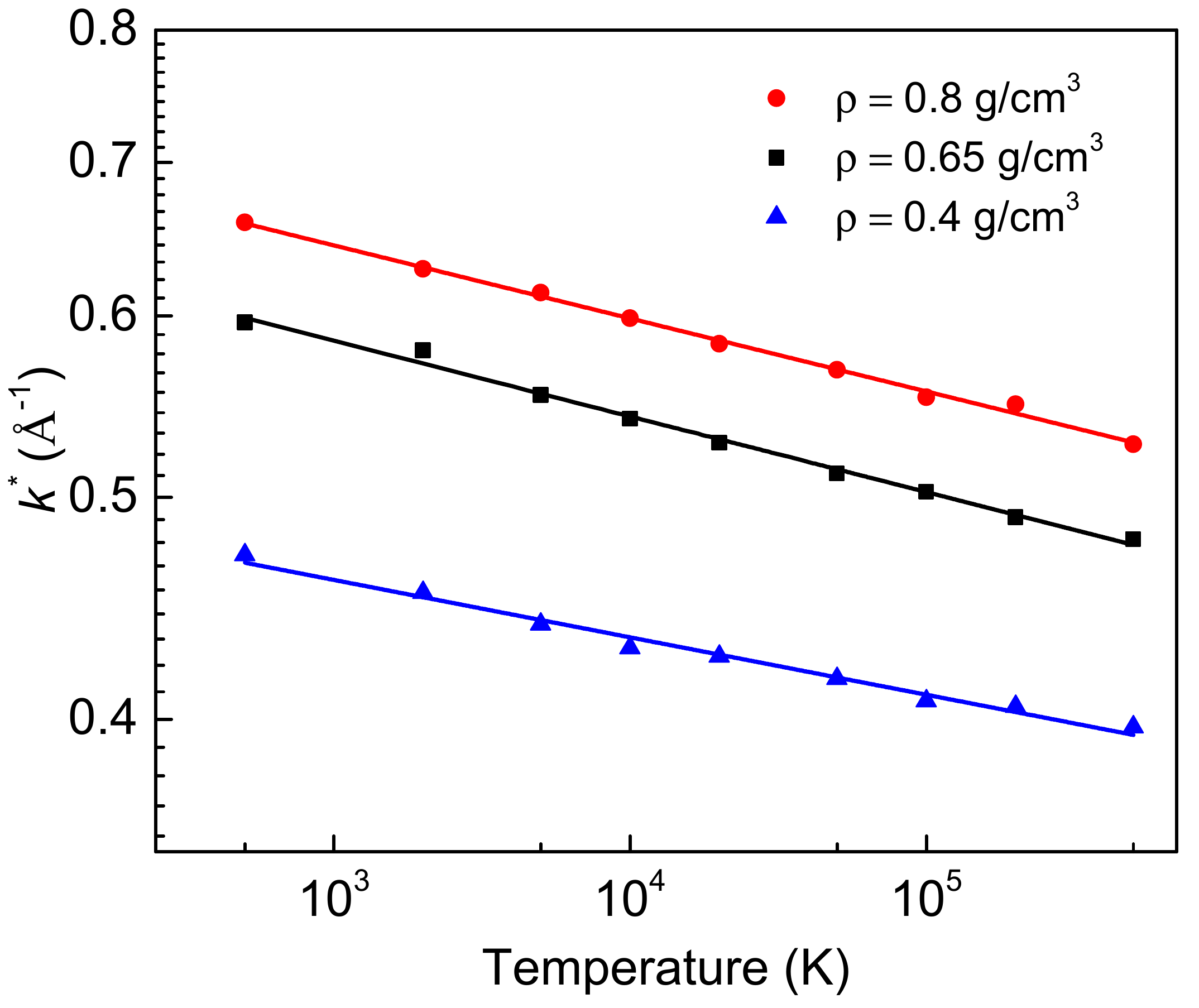}}}
\end{center}
\caption{Dependence of $k^\star$ on temperature. The lines are fits to the power-law dependence.}
\label{kt}
\end{figure}

We plot $k^\star$ as a function of temperature in double-logarithmic plot in Fig. \ref{kt}. Fig. \ref{kt} shows that $k^\star$ and $\lambda$ obey the power law

\begin{equation}
\begin{split}
&k^\star\propto\frac{1}{T^\alpha}\\
&\lambda\propto T^\alpha
\end{split}
\label{lambda1}
\end{equation}

\noindent where $\alpha$ is about $0.1$.

The existence of a power law is important because it often implies an underlying dynamical process leading to universal scaling relationships, as it does in, for example, the phase transitions area. We will re-visit this point below in our evaluation of $l$.

As proposed above, the crossovers in Fig. \ref{dispersion} are related to the shortest wavelength of the system governed by the mean free path $l$. $l$ can be directly evaluated from viscosity in the mean-free path regime: $\eta=\frac{1}{3}\rho{\bar v}l$, where $\rho$ is density and ${\bar v}$ is average velocity. Using experimental $\eta$ from the NIST database \cite{NIST} at 500 K, we calculate $l$ and the corresponding $k$-point as $\frac{\pi}{l}$. We plot the calculated $k$ points as red stars in Fig. \ref{dispersion} and observe that they lie close to both $k^\star$ calculated on the basis of Eq. (\ref{lambda}) and to the crossover of dispersion curves. The NIST database does not extend to high temperature, and so we fit the experimental low-temperature viscosity to the power law (see the discussion below) and extrapolate $\eta$ to high temperature. Similarly to 500 K, the calculated $k$-points from extrapolated viscosity remain close to $k^\star$ calculated on the basis of Eq. (\ref{lambda}) and to the crossover of dispersion curves. The increase of the distance between red and blue stars at high temperature compared to low can be related to the reduced reliability of extrapolation to high temperature.

The proximity of $k$-points evaluated from $\lambda$ in Eq. (\ref{lambda}) and the mean free path $l$ to each other and to the crossover of dispersion curves constitutes our {\it second} quantitative evidence in support to our hypothesis that the crossover of dispersion curves is related to the crossover from the phonon regime to the MFP regime where the mean-free path limits the shortest available wavelength as $\lambda=l$.

Importantly, the small value of $\alpha$ we observe for $\lambda$ in Eq. (\ref{lambda1}) is the same as the temperature exponent of the mean free path $l$ in both experiment and theory. Experimentally, evaluating $l$ from the experimental supercritical gas-like viscosity gives $l\propto{T^\alpha}$ with $\alpha$ close to 0.1 \cite{natcom}.

Interestingly, the same result $l\propto{T^\alpha}$ with $\alpha$ close to 0.1 follows from the kinetic gas theory. Approximating the Enskog series by the first term and considering the interatomic interaction in the form of the inverse-power law $U\propto\frac{1}{r^m}$ gives viscosity $\eta$ as \cite{chapman}

\begin{equation}
\begin{split}
&\eta\propto T^s\\
&s=\frac{1}{2}+\frac{2}{m-1}
\end{split}
\label{exp}
\end{equation}

The temperature dependence of $l$ can be predicted using Eq. (\ref{exp}) and viscosity in the MFP regime $\eta=\frac{1}{3}\rho{\bar v}l$ and noting that ${\bar v}\propto T^{0.5}$. This gives $l\propto T^{s-0.5}$.

In the limit of large $m$ corresponding to the hard sphere system, (\ref{exp}) predicts $s=\frac{1}{2}$ and, therefore, temperature-independent $l$. In this limit, $l$ is governed by density only. Some care is needed to evaluate $s$ for the LJ potential used in this work. We recall that the function describing the repulsive part of the LJ potential is governed by both repulsive $\frac{1}{r^{12}}$ and attractive $\frac{1}{r^6}$ terms. The net result is that the effective repulsive function varies as $U\propto\frac{1}{r^m}$ with $m=18$--$20$ \cite{dyre}. Using this $m$ in (\ref{exp}) gives $s\approx 0.6$. Using $l\propto T^{s-0.5}$ gives $l\propto T^{0.1}$, and we obtain the same exponent calculated for $\lambda$ in (\ref{lambda1}).

The coincidence of the temperature behavior of $\lambda$ and $l$ as power law $\propto T^\alpha$ with the same exponent $\alpha$ constitutes our {\it third} quantitative evidence in support to our hypothesis that the crossover of dispersion curves is related to the crossover from the sound to the MFP regime and $\lambda=l$.

Taken together, the three quantitative estimates support the interpretation of the crossover to the MFP regime, rather than a trivial deviation of dispersion curves from linearity close to the zone boundary.

\begin{figure}
\begin{center}
{\scalebox{0.33}{\includegraphics{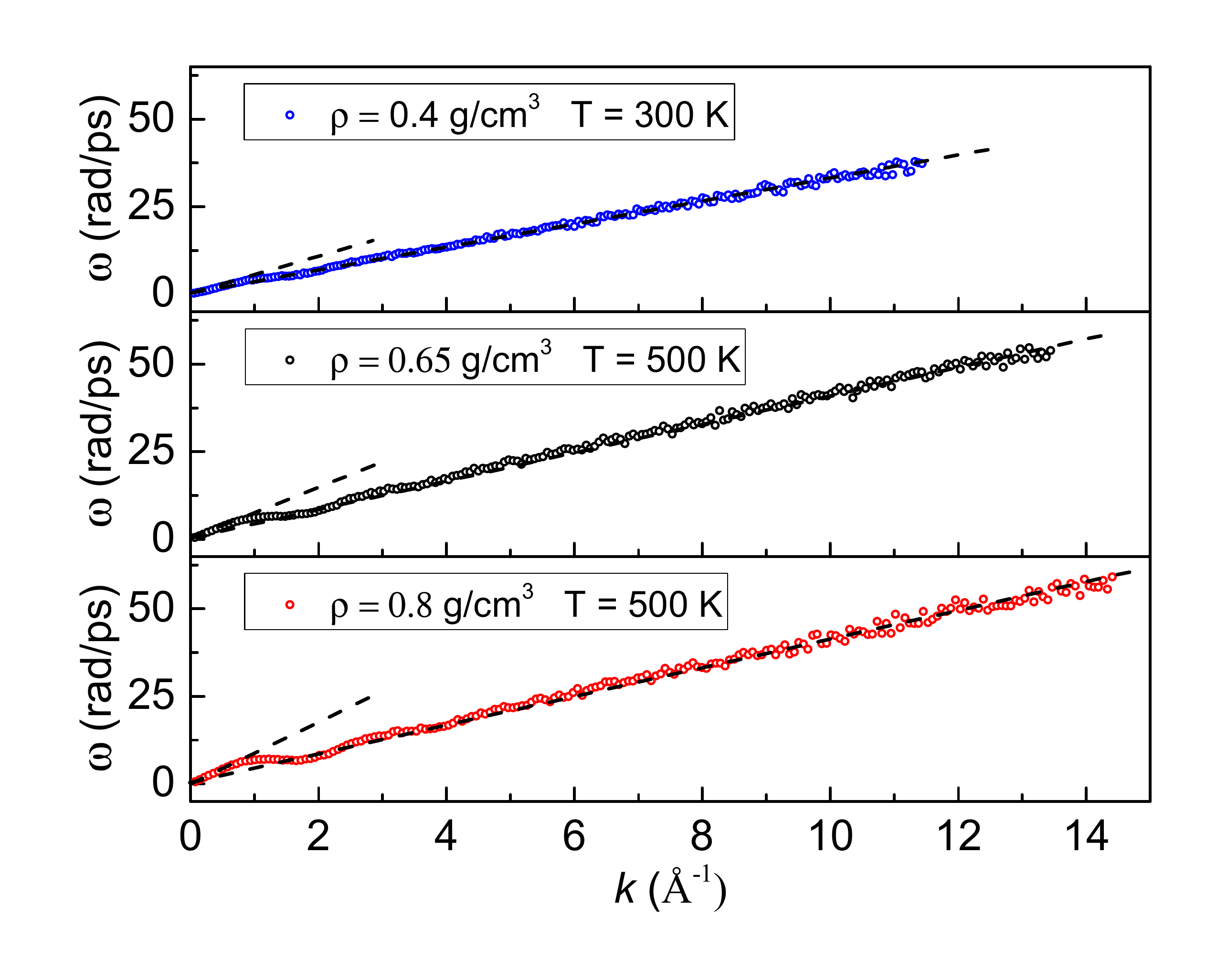}}}
\end{center}
\caption{Phonon dispersion curves of supercritical Ar at large $k$ at different density and temperature. Dashed lines are guides for the eye and show the speed of sound at small $k$ and thermal velocity at large $k$.}
\label{free}
\end{figure}

We note that at larger $k$, the mean free path regime is followed by the free-particle regime. Indeed, very large $k$ and $\omega$ correspond to particles moving short distances at short times, i.e. the regime of free particles and the dispersion $\omega=vk$, where $v$ is particle velocity (this also applies to crystals where large $k$ correspond to scattering from both free particles and phonons in higher Brillouin zones). We have calculated the dispersion relationship for large $\omega$ and $k$ and show examples in Fig. \ref{free}. The calculated slope at large $k$ agrees with the most probable thermal speed $v_p=\sqrt{2k_{\rm B}T/m}$ within 9--13\% for different densities. Hence, the MFP regime we considered earlier is intermediate between the sound regime and the free-particle regime in terms of $k$-values.

We make a remark regarding the values of $l$. The shortest wavelengths at $k$-points in Fig. \ref{kt} and the corresponding $l$ are approximately 10--15 \AA, or about 3--4 interatomic separations (the small range of $l$ is due to its slow decrease on the isochores as discussed earlier). This is shorter than $l$ generally envisaged in the kinetic theory of gases and is to be expected for dense supercritical fluids as compared to dilute gases \cite{chapman}. This can explain why the slope of $\omega$ vs $k$ differs from thermal velocity in the MFP regime (thermal velocity is recovered at large $k$ as discussed above). 

Even though $l$ is short, it is nevertheless important from the thermodynamic point of view because it cuts off phonons with largest frequencies which contribute most to the system energy \cite{ropp}. As a result, $c_v$ decreases from about 2 at the Frenkel line (this value has contributions from the kinetic term $\frac{3}{2}$ and the potential term of the longitudinal mode $\frac{1}{2}$) to the value close to $c_v$ of the ideal gas, $\frac{3}{2}$. This is seen in Fig. \ref{cv}. Short $l$ also implies that, since $l$ is an average property, its relative fluctuations around the mean value may be non-negligible, which could explain that in some cases $\lambda$ and $l$ may deviate from the observed crossover of dispersion curves in Fig. \ref{dispersion}.

We note that high temperatures were required to detect the decreasing $k$ at the crossover at fixed density. Although these temperatures might seem unusually high, there are three reasons why they are relevant to real systems and experiments. First, liquid argon remains an unmodified system up to fairly high temperature: the first ionization potential of condensed liquids is on the order of 10$^5$ K. Hence most of our temperature range where the dispersion relations and crossovers are seen corresponds to the unmodified non-ionized argon describable by the LJ potential. Second, performing experiments at realistic constant pressure, rather than constant density used here, lowers the crossover temperature significantly due to faster increase of the mean free path when volume increases. Third, performing the experiments in systems with lower critical point such as Ne implies that the crossover temperature is lower: the crossover at the largest temperature simulated here is predicted to be lower by the ratio of Ar and Ne critical temperatures, or over 3 times.

\section{Summary}

In summary, we presented evidence for propagating solid-like longitudinal phonons deeply in the supercritical regime, with wavelengths extending to interatomic separations and observed the crossover of dispersion curves. By studying temperature dependence of the shortest available wavelength and mean free paths, we related this effect to the crossover from the collective phonon to the collisional mean-free path regime of particle dynamics.

\section{Acknowledgments}

This research utilised MidPlus computational facilities supported by QMUL Research-IT and funded by the EPSRC Grant EP/K000128/1. We acknowledge the support of the Royal Society, RFBR (15-52-10003) and CSC.

\section{Author contributions}

L.W. and K.T. wrote the main manuscript text and prepared figures 1-7. L.W., C.Y., M.T.D., A.V.M., V.V.B. and K.T. reviewed the manuscript and have contributed equally to this work.

\section{Additional information}
Competing Interests:  The authors declare no competing interests.


\begin{thebibliography}{99}

\bibitem{landau} L. D. Landau and E. M. Lifshitz, Statistical Physics, (Pergamon Press, 1969).

\bibitem{burkel} E. Burkel, Rep. Prog. Phys. {\bf 63}, 171 (2000).

\bibitem{pilgrim2} W. C. Pilgrim and C. Morkel J. Phys.: Condens. Matter. {\bf 18}, R585 (2006).

\bibitem{rec-review} W. C. Pilgrim and C. Morkel, J. Phys.: Cond. Matt. {\bf 18}, R585 (2006).

\bibitem{hoso} S. Hosokawa et al Phys. Rev. Lett. {\bf 102}, 105502 (2009).

\bibitem{hoso3} S. Hosokawa, M. Inui, Y. Kajihara, S. Tsutsui and A. Q. R. Baron, J. Phys.: Condens. Matt. {\bf 27}, 194104 (2015).

\bibitem{mon-na} V. M. Giordano and G. Monaco, PNAS {\bf 107}, 21985 (2010).

\bibitem{mon-ga} V. M. Giordano and G. Monaco, Phys. Rev. B {\bf 84}, 052201 (2011).

\bibitem{sn} S. Hosokawa et al, J. Phys.: Condens. Matt. {\bf 25}, 112101 (2013).

\bibitem{hydro} L. D. Landau and E. M. Lifshitz, Fluid Mechanics (Butterworth-Heinemann, 1987).

\bibitem{ropp} K. Trachenko and V. V. Brazhkin, Rep. Prog. Phys. {\bf 79}, 016502 (2016).

\bibitem{chapman} S. Chapman and T. G. Cowling, The Mathematical Theory of Non-uniform gases (Cambridge University Press, 1995).

\bibitem{deben} E. Kiran, P. G. Debenedetti and C. J. Peters, Supercritical Fluids: Fundamentals and Applications, NATO Science Series E: Applied Sciences, {\bf 366} (Kluwer Academic Publishers, Boston, 2000).

\bibitem{pre} V. V. Brazhkin, Yu. D. Fomin, A. G. Lyapin, V. N. Ryzhov and K. Trachenko, Phys. Rev. E {\bf 85}, 031203 (2012).

\bibitem{prl} V. V. Brazhkin, Yu. D. Fomin, A. G. Lyapin, V. N. Ryzhov, E. N. Tsiok and K. Trachenko, Phys. Rev. Lett. {\bf 111}, 145901 (2013).

\bibitem{phystoday} V. V. Brazhkin and K. Trachenko, Physics Today {\bf 65(11)}, 68 (2012).

\bibitem{pastore1} R. Pastore, G. Pesce, A. Sasso, and M. P. Ciamarra, J. Phys. Chem. Lett. {\bf 8}, 1562 (2017).

\bibitem{pastore2} R. Pastore, G. Pesce and M. Caggioni, Sci. Rep. {\bf 7}, 43496 (2017).

\bibitem{yang} C. Yang, M. T. Dove, V. V. Brazhkin, and K. Trachenko, Phys. Rev. Lett. {\bf 118}, 215502 (2017).

\bibitem{jpcm} Yu. D. Fomin et al, J. Phys.: Condens. Matt. 28, 43LT01 (2016).

\bibitem{jpcm1} Yu. D. Fomin et al, J. Phys.: Condens. Matt. 30, 134003 (2018).

\bibitem{natcom} D. Bolmatov, V. V. Brazhkin and K. Trachenko, Nat. Comm. {\bf 4}, 2331 (2013).

\bibitem{md} I. T. Todorov, B. Smith, M. T. Dove, and K. Trachenko, J. Mater. Chem. {\bf 16}, 1911 (2006).

\bibitem{ling} L. Wang et al, Phys. Rev. E {\bf 95}, 032116 (2017).

\bibitem{monaco} V. M. Giordano and G. Monaco, PNAS {\bf 107}, 21985 (2010).

\bibitem{Balucani} U. Balucani and M. Zoppi, Dynamics of the Liquid State, (Oxford University Press, New York, 1995).

\bibitem{giordano-PRB} V. M. Giordano and G. Monaco, Phys. Rev. B {\bf 84}, 052201 (2011).

\bibitem{stas} S. O. Yurchenko et al, J. Chem. Phys. {\bf 148}, 134508 (2018).

\bibitem{dyre} N. P. Bailey, U. R. Pedersen, N. Gnan, T. B. Schr{\o}der and J. C. Dyre, J. Chem. Phys. {\bf 129}, 184507 (2008)

\bibitem{NIST} NIST database Thermophysical properties of fluid systems, see http://webbook.nist.gov/chemistry/fluid

\end{thebibliography}
\end{document}